\documentclass[11pt,oneside,reqno]{amsart}

\usepackage{graphicx}
\usepackage{pict2e}
\usepackage{amssymb}
\usepackage{amsthm}                % better theorem 
\usepackage[margin=1in]{geometry}  % set the margins to 1in on all sides
\usepackage{graphics}
\usepackage{epstopdf}
\usepackage{amsmath}
\usepackage{graphicx}
\usepackage{subfigure}
\usepackage{latexsym}
\usepackage{amsmath}
\usepackage{amsfonts}
\usepackage{amssymb}
\usepackage{tikz}
\usepackage{mathdots}
\usepackage{comment}
\usepackage{float}
\usepackage[mathscr]{euscript}
\usepackage{enumerate}
\usepackage{epsfig}
\usepackage { graphicx }
\usepackage { hyperref }
\usepackage { graphics }
\usepackage { graphicx }
\usepackage{booktabs}
\usepackage{wrapfig,lipsum,booktabs}

\usepackage{caption}

\usepackage{algorithm}

\usepackage{etoolbox}

\newcommand{\Xspace}        {{\mathbb X}}

\newcommand {\mm}[1] {\ifmmode{#1}\else{\mbox{\(#1\)}}\fi}

\newcommand{\Hgroup}        {{\sf H}}

\title{ Inferring Quality in Point Cloud-based 3D Printed Objects using Topological Data Analysis}
\author{Paul Rosen}
\address{University of South Florida}
\email{prosen@usf.edu}

\author{Mustafa Hajij}
\address{University of South Florida}
\email{mhajij@usf.edu}

\author{Junyi Tu}
\address{University of South Florida}
\email{junyi@mail.usf.edu}

\author{Tanvirul Arafin}
\address{University of South Florida}
\email{tanvirul@mail.usf.edu }

\author{Les Piegl}
\address{University of South Florida}
\email{lespiegl@usf.edu }

\begin{document}

% #################################################
% Sets up the beginning of the document (do not modify) 
% #################################################
\maketitle

% ############################################
% Here is the fun part. Please enter your:            
%      Name (First, Middle initial, Last)                  
%      Your ORCID number replace the "000-0000-1234-5678" part.    
% ############################################

%\authorSection{
%	\anAuthor{Paul Rosen}{0000-0002-0873-9518}{1},
%	\anAuthor{Mustafa Hajij}{}{2}, 
%	\anAuthor{Junyi Tu}{0000-0001-7026-7454}{3},
%	\anAuthor{Tanvirul Arafin}{}{4},
%	\anAuthor{Les Piegl}{0000-0003-0629-8496}{5}
%}

% ######################################
% Here we need your affiliation and contact e-mail. Please edit   
% affiliation as well as the e-mail fields.                                  
% ######################################
%\affiliationSection{
%	\anAffiliation{1}{University of South Florida}{prosen@usf.edu}
%	\anAffiliation{2}{University of South Florida}{mhajij@usf.edu}
%	\anAffiliation{3}{University of South Florida}{junyi@mail.usf.edu}
%	\anAffiliation{4}{University of South Florida}{tanvirul@mail.usf.edu}
%	\anAffiliation{5}{University of South Florida}{lespiegl@usf.edu}
%}

% ######################################
% Please decide on who the corresponding author is going to be 
% and complete the section below                                            
% ######################################
%\correspondingAuthor{Paul Rosen}{prosen@usf.edu}

% ########################################
% Please type in your abstract below after the "\abstract{" part.
% ########################################
\begin{abstract}
Assessing the quality of 3D printed models before they are printed remains a challenging problem, particularly when considering point cloud-based models. This paper introduces an approach to quality assessment, which uses techniques from the field of Topological Data Analysis (TDA) to compute a topological abstraction of the eventual printed model. Two main tools of TDA, Mapper and persistent homology, are used to analyze both the printed space and empty space created by the model. This abstraction enables investigating certain qualities of the model, with respect to print quality, and identifies potential anomalies that may appear in the final product.
\end{abstract}

% ########################################
% Please choose at least 3-5 good keywords and list them after the
% "\noindent \textbf{Keywords:}" part.
%
% The DOI part will be edited by us so please do not change that.
% ########################################

%\doi{10.3722/cadaps.2018.xxx-yyy}

% ################################
% You may now start your paper with an introduction
% ################################

\section{Introduction}
3D printing is gaining incredible popularity in low-yield manufacturing for customized or specialized parts. However, assessing the quality of models before they are printed remains a challenging problem~\cite{telea2011voxel}, particularly when you consider point cloud-based models~\cite{oropallo2018point}, such as those that come from 3D scanners. This paper introduces an approach to quality assessment, which uses techniques from the field of Topological Data Analysis (TDA) to compute a topological abstraction of the eventual printed model and the empty space around and contained within it. This abstraction enables investigating certain properties of the model, with respect to print quality, and identifies potential anomalies that may appear in the final product.

\section{Mapper and Persistent Homology}

\begin{figure}[!h]
\centering
%\vspace{-5pt}
\includegraphics[width=0.975\linewidth]{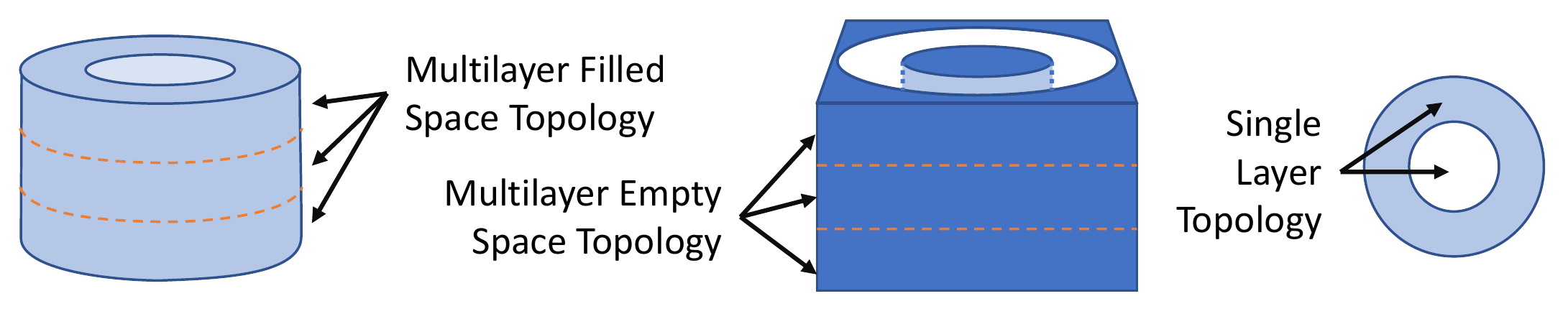} \\
\caption{Our approach uses Mapper to look at the filled space topology of multiple layers (left) and empty space topology of multiple layers (middle). It uses persistent homology to understand the topology of a single layer (right).}
\label{fig.topo_example}
\end{figure}

This approach uses 2 of the fundamental tools of TDA, namely Mapper~\cite{singh2007topological} and persistent homology~\cite{edelsbrunner2000topological}, 
to provide users with feedback about 
their models (see Figure~\ref{fig.topo_example}). Mapper is used in 2 ways. First, it is used to extract information about the layer-by-layer connectivity of the model to be printed, providing an abstraction of the overall shape of the object. Second, it is used to determine the topology of the empty space contained within and surrounding the printed model. Persistent homology on the other hand is a tool that normally is used to provide a multiscale view of connected components, holes/tunnels, and voids in data of any dimension. Our approach uses persistent homology for the detection of connected components and holes within a printer layer.

The inner workings and associated details of both Mapper and persistent homology are quite complicated, and so we refer the reader to prior work for a better understanding~\cite{edelsbrunner2000topological, singh2007topological}. We will instead provide an intuition about the types of structures captured by each of these tools.

\subsection{Mapper}
Mapper is a TDA tool that provides a graph-based abstraction of the topology of a mesh or point-based data. Mapper construction starts by first parameterizing and slicing the data. In our case the parameterization is vertical. 

The graph vertices are created from connected components identified within each layer. In other words, the connected components of the layer are ``collapsed'' into graph vertices. There are many variations on identifying connected components from points. We use the persistent homology approach, introduced in the next subsection. 

Finally, graph edges are added between components that touch on neighboring layers. This connection is made by adding a small amount of overlap to each layer. If one or more points in the overlap region are contained within connected components from 2 different layers, those component vertices receive a graph edge. The resulting graph can describe the overall topology of the connected components of a printed object.

%
%\begin{wrapfigure}[14]{r}{0.6\linewidth}
\begin{figure}[!h]
\centering
%\vspace{5pt}
\includegraphics[width=0.95\linewidth]{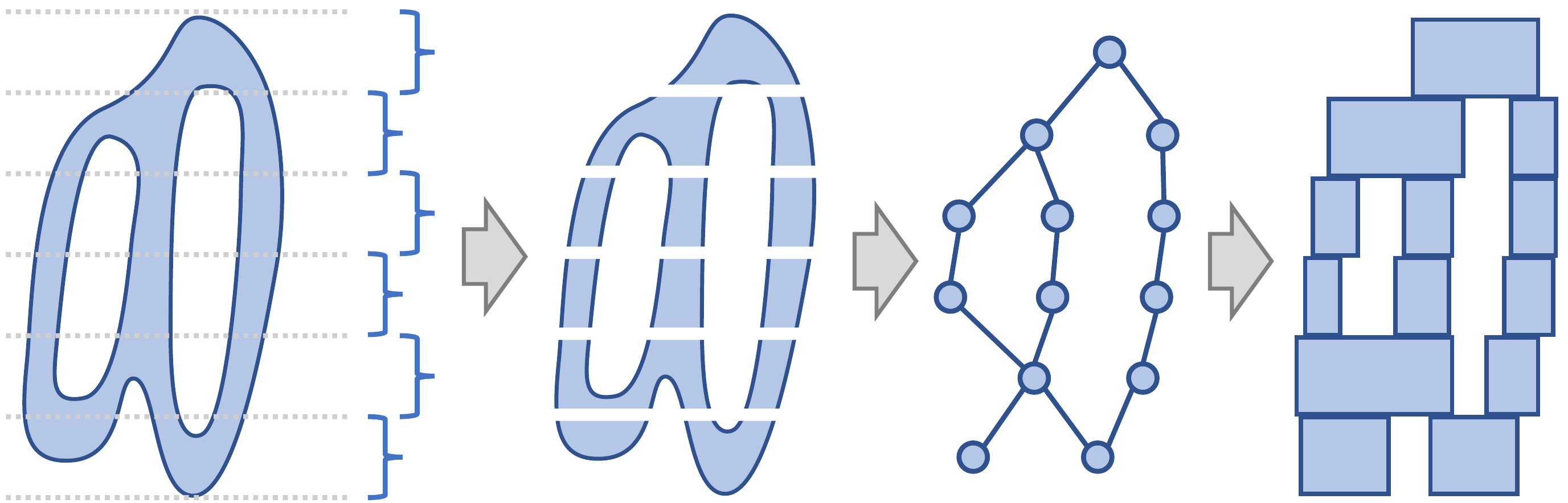}\\
\vspace{-10pt}
\subfigure[]{ }\hspace{125pt}
\subfigure[]{ }\hspace{100pt}
\subfigure[]{ }\hspace{90pt}
\subfigure[]{ }\hspace{0pt}
\vspace{-5pt}
\caption{Example of Mapper on a mesh. The (a) model is (b) sliced. (c) Connected components are collapsed to vertices and edges added for components that touch. (d) Finally, an illustration of the printed object is shown.}
\label{fig.mapper_example}
\end{figure}
%\end{wrapfigure}
%

Figure~\ref{fig.mapper_example} shows an example of Mapper on a simple domain. First, (a) the input model is (b) sliced with layer thickness being set to equal the 3D printer's layer resolution. Next, (c) the connected components are found and edges added when they touch. (d) Finally, the illustration of the printed object is shown for comparison. The nodes of the Mapper graphs do not provide any insight into the size or shape of a given connected component. Instead they provide insight into which components touch and how those components may or may not form holes in the output model.

\begin{figure}[!h]
\centering
%\vspace{5pt}
\includegraphics[width=0.90\linewidth]{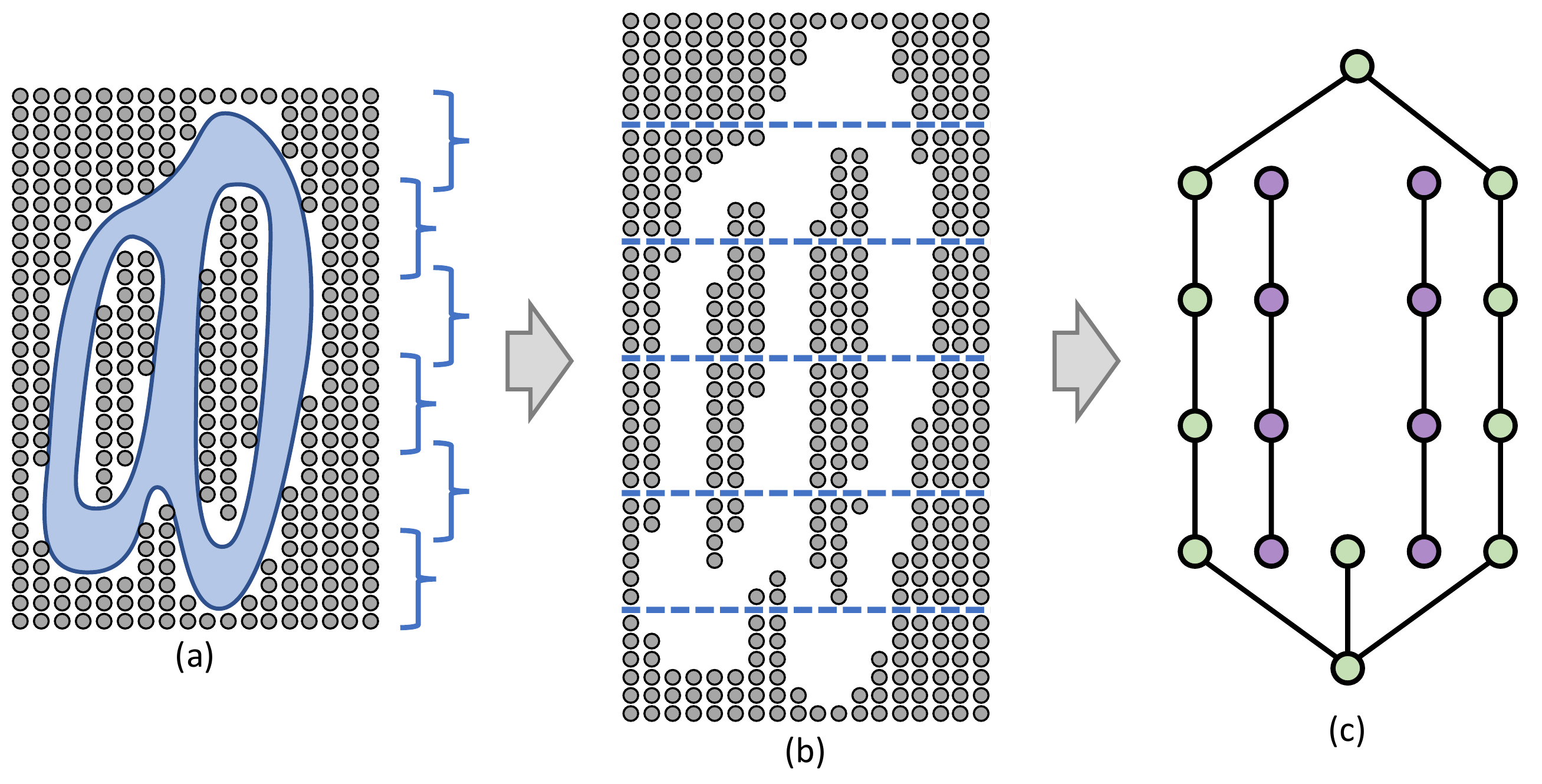}
\caption{Example of Mapper on the empty space of a mesh. The (a) model has its empty space filled with points and is (b) sliced. (c) The connected components are collapsed to vertices and edges added for components that overlap. For illustration purposes, the vertices here are colored green for outside and purple for inside the model.}
\label{fig.mapper_empty}
\end{figure}

Calculating the Mapper graph on the empty space is a similar process. However, to calculate the graph, the empty space first needs to be filled. This is done by populating the empty space with points. Then, Mapper construction proceeds identically on the empty space points. The approach is illustrated in Figure~\ref{fig.mapper_empty}.

The calculation of Mapper is relatively inexpensive. The slicing operation is linear in the number of points. The connected component detection is naively quadratic in the number of points per layer, but this can be improved with spatial partitioning. The overall performance can be improved by using a parallelized algorithm~\cite{hajij2017distributed}.

\begin{comment}
%
\begin{wrapfigure}[9]{r}{0.25\linewidth}
\includegraphics[width=0.975\linewidth]{Figures/torus.pdf}
\caption{A 3D torus.}
\label{fig.torus}
\end{wrapfigure}
%
\end{comment}

\subsection{Persistent Homology}
\begin{figure}[!h]
\centering
\subfigure[\label{fig.annulus}]{\includegraphics[width=0.225\linewidth]{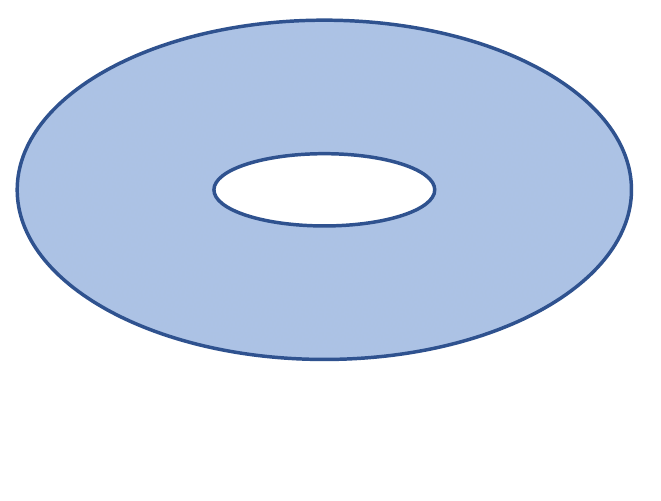}}
\hfill
\includegraphics[width=0.74\linewidth]{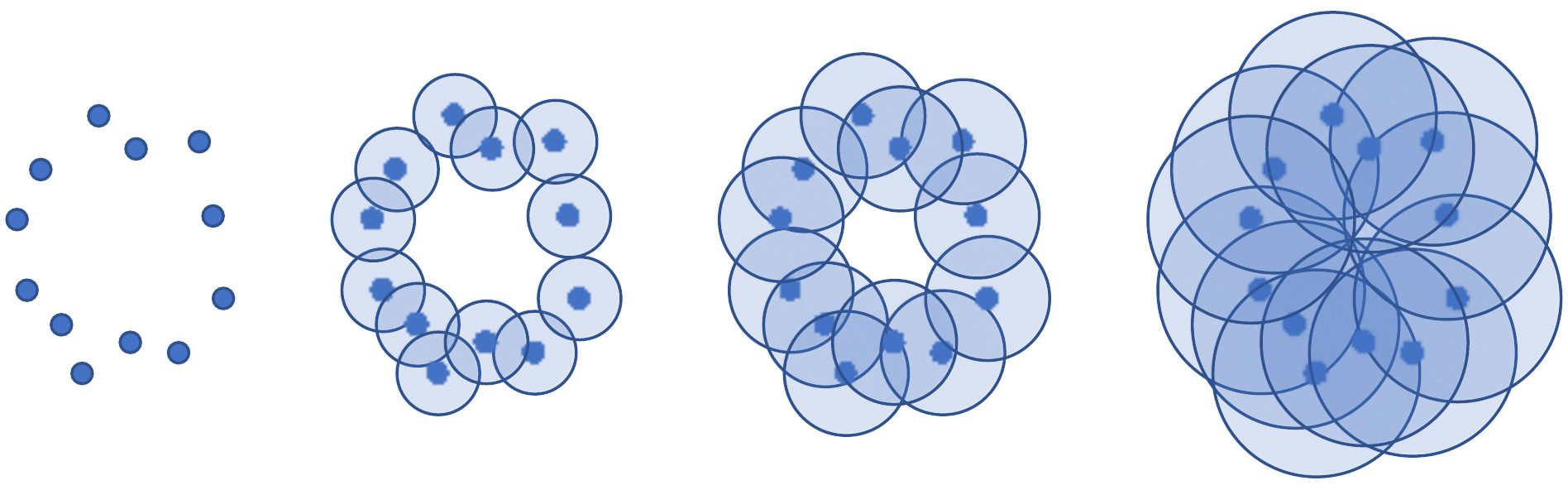}\\
\vspace{-27pt}
\hspace{120pt}
\subfigure[]{}\hspace{75pt}
\subfigure[]{}\hspace{85pt}
\subfigure[]{}\hspace{100pt}
\subfigure[]{}\hspace{35pt}
\caption{(a) An annulus. (b-e) Example of persistent homology as it relates to a point-based annulus. As points are thickened, from (b) to (e), a hole/tunnel forms in (c) and closes in (e).}
\label{fig.ph}
\end{figure}

Given a topological space $\Xspace$, the homology groups $\Hgroup_0(\Xspace)$, $\Hgroup_1(\Xspace)$, and $\Hgroup_2(\Xspace)$, describe the connected components, holes/tunnels, and voids of the space, respectively. For example, consider the annulus 
in Figure~\ref{fig.annulus}. It has a 
single connected component. It also has a single hole/tunnel through the middle. Finally, it contains no void.

%\vspace{-12pt}
%
%\begin{wrapfigure}[11]{r}{0.525\linewidth}

The multiscale notion of homology, called persistent homology, extracts the homology groups of a set of points considering different resolutions.  A topological feature therefore has a minimum resolution where it first appears, known as the birth time, and a maximum resolution it is still visible, known as its death time. This can be intuitively thought of as the thickening of points. Figure~\ref{fig.ph}(b-e) shows an example. Starting with (b) 12 points, the points are thickened, until (c) they form a single connected component with a hole. As the points continue to thicken (d) the hole remains visible, until (e) the thickness of the points closes it.

The performance calculating $\Hgroup_0$ connected components is the same process per layer as with Mapper, naively quadratic. Finding the $\Hgroup_1$ homology groups (i.e.\ holes/tunnels) in persistent homology is quite expensive. This calculation builds a simplicial complex on the data in the form of a boundary matrix and performs a reduction, similar to Gaussian elimination, which leads to a worst case performance that is cubic in the number of points. The average run time is linear with a large time constant. We mitigate this by pre-extracting per-layer connected components and running this calculation only on those components. 

%Given a set of points where the scale of features is unknown, the multiscale notion of homology, known as persistent homology, extracts the structure of the data by considering all potential feature scales. Given a set of points, the scale at which a feature first appears is known as its birth, while the largest scale at which it appears is known as its death.

\subsection{Link Between Mapper and Persistent Homology}
The most direct link between Mapper and persistent homology is to use the persistent homology approach in the calculation of $\Hgroup_0(\Xspace)$ homology groups (i.e.\ connected components) for the individual slices of the Mapper algorithm. However, we augment the conventional Mapper implementation by further attaching the $\Hgroup_1(\Xspace)$ homology groups (i.e.\ holes/tunnels) to the individual nodes of the Mapper graph. By doing this, the number of holes in each connected component is retained for further analysis.

\section{The Topology of 3D Printing}
It turns out that both Mapper and persistent homology have direct applications to 3D printing anomaly detection. For Mapper, the slicing operation has a direct corollary in the layers of a 3D printer. Therefore, the slice thickness, known as the cover, can be set to the same value as the thickness of a single layer on the 3D printer (i.e.\ the z resolution). For persistent homology, the calculation of connected components is the same as a physically connected components within a single layer. The holes within each layer represent the holes within the model. These can be determined by targeting the xy resolution of 3D printer of interest. Furthermore, using the empty space, Mapper can provide information about the watertightness of the model.

\subsection{Visualization}
\begin{figure}[!h]
\centering
\includegraphics[width=0.975\linewidth]{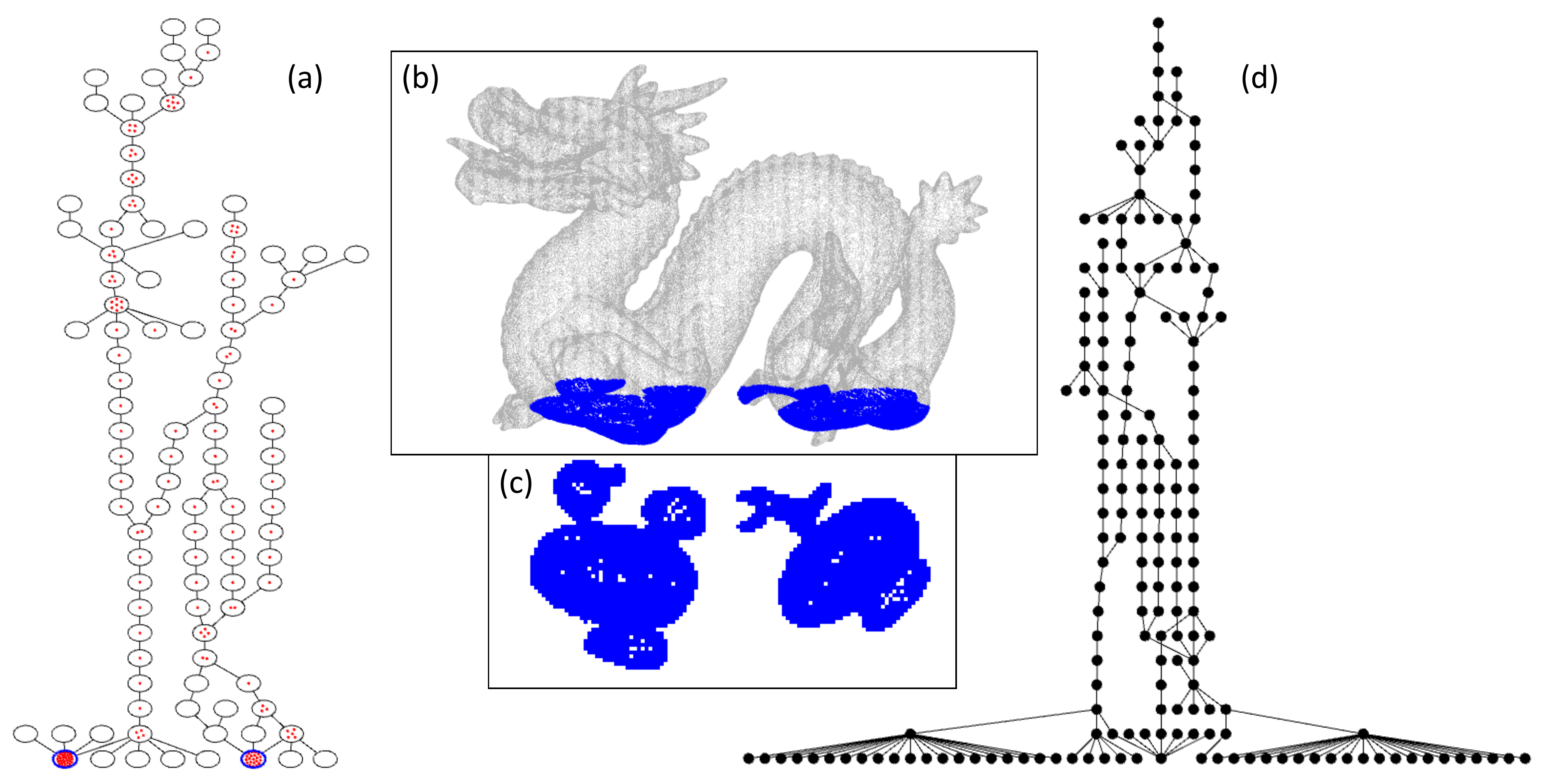}
\caption{Our software with the Stanford Dragon dataset. (a) The filled space topology is shown as a Mapper graph with holes denoted as red dots. (b) A 3D view and (c) a single slice view are shown for detail. (d) The empty space topology is shown only as the Mapper graph.}
\label{fig.software}
\end{figure}

Once the topology of the point cloud has been calculated, we provide a visualization for inspecting the data. The visualization contains 4 components. The first, and most important, is the Mapper graph of the printed model, as seen in Figure~\ref{fig.software}(a). The Mapper graph nodes shows the individual connected components of the model. In addition, each tunnel going through the connected component is represented by a red point in the node visualization. The next visualization, as seen in Figure~\ref{fig.software}(d) is the Mapper graph calculated on the empty space of the model, instead of the filled space. The last 2 visualizations are: the 3D point cloud (Figure~\ref{fig.software}(b)), with regions highlighted based upon the selection of Mapper graph nodes, and a 2D slice visualization (Figure~\ref{fig.software}(c)), again based upon nodes selected in the Mapper graph.

\begin{figure}[!t]
\centering
\includegraphics[width=0.975\linewidth]{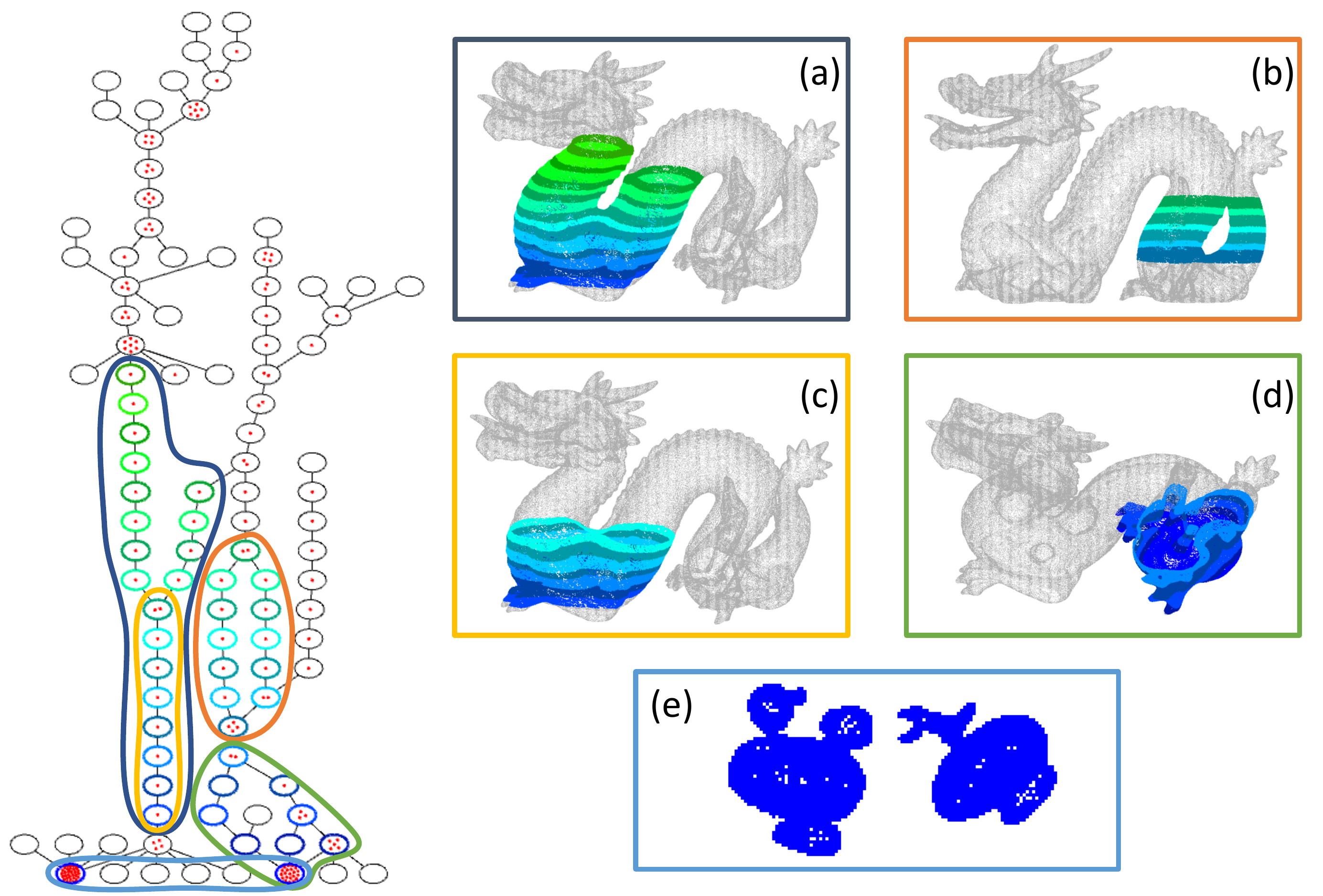}
\caption{Results of Dragon dataset. The Mapper graph of the filled space (left) has 5 different portions (a-e) highlighted (right).}
\label{fig.results}
\end{figure}

\section{Results}
We implemented our approached using a number of tools. First, data is converted into a point by any method of choice, such as \cite{oropallo2018point}. In our case, PLY or STL files had their vertices extract directly. Our Mapper implementation is in Java. The software loads a point cloud, slices it, detects connected components, and exports the Mapper graph and connected component points for both the filled space and empty space. Each filled space connected component is then fed into Ripser\footnote{Ripser: \url{https://github.com/Ripser/ripser}} for persistent homology detection of holes/tunnels. For the visualization of the Mapper graph, the layout was calculated using Graphviz\footnote{Graphvis: \url{https://www.graphviz.org/}}. The data was then fed into our visualization tool built using Processing\footnote{Processing: \url{https://processing.org/}}.

We tested our approach on the Dragon dataset from the Stanford 3D Scanning Repository. We used the points from the reconstructed dataset, which contained approximately 437,000 points. The question we were after was, if someone was to try to rasterize these points directly for 3D printing (ignoring any mesh connectivity), what sort of anomalies would occur. We first scaled the model to a height of 10 cm. We then chose the z resolution to be 3.3 mm and xy resolution to be 1.0 mm.

\subsection{Original Model}

After running our pipeline, the results are displayed in Figure~\ref{fig.software} and Figure~\ref{fig.results}. In Figure~\ref{fig.results}, the tree on the left overviews the entire structure of the graph. We will concentrate on the few circled regions. 

First, starting with Figure~\ref{fig.results}(c) in yellow, notice that this region represents a portion of the body of the dragon. In this region, each ring forms a single connected component, each with a single hole through the middle. That is until the topmost ring, where a single connected component has 2 holes, beginning the bifurcation of the upper front and middle portions of the body, as seen in Figure~\ref{fig.results}(a) in dark blue. This feature can be observed in the graph by looking at the top most node in the yellow circle. Notice 2 red dots, indicating 2 holes in that component.

Next, notice the region Figure~\ref{fig.results}(b) in orange. In this region, the model itself splits and comes back together leaving a hole between the torso and tail. This can be observed in the graph as well. Starting after the bottom node of orange region, the graph bifurcates, indicating a split in the connected components, and merges again at the top. This splitting and merging pattern is indicative of an exterior hole in the model. This same type of splitting and merging behavior can also be noticed in the graph region circled in green and associated with Figure~\ref{fig.results}(d). This hole is caused by the leg and body coming together. However, it is difficult to observe by looking at the 3D imagery of the point cloud. In fact, we could not find a good viewing angle that showed this hole directly.

\begin{figure}[!b]
\centering
\vspace{-10pt}

\subfigure[\label{fig.corrected_results.printed}]{\includegraphics[height=4.5in]{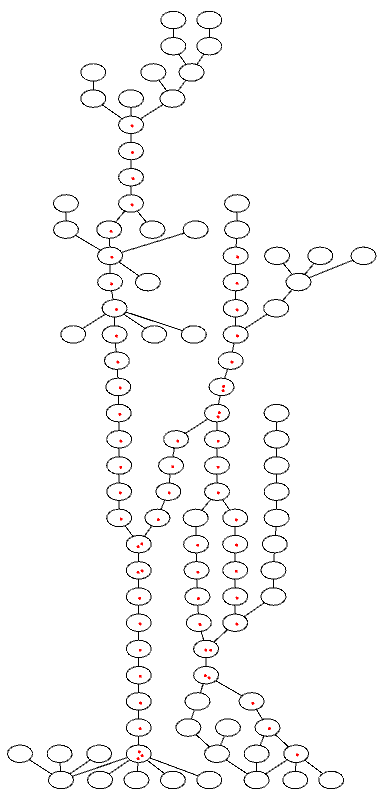}}
\hspace{50pt}
\subfigure[\label{fig.corrected_results.empty}]{\includegraphics[height=4.5in]{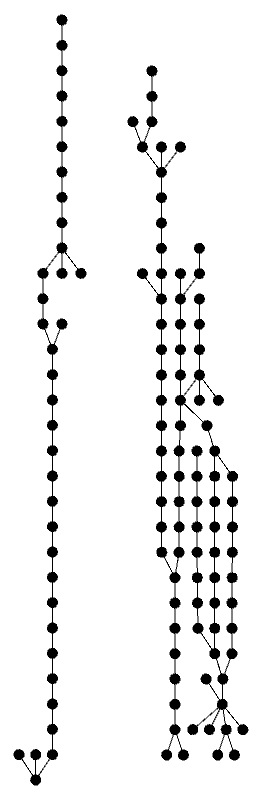}}

\caption{Error Corrected Results of Dragon dataset. (a) The filled space shows a single connected component and holes only on the interior. (b) The empty space has 2 connected components, (left) the outside of the model and (right) the inside of the model. This indicates that the model is now watertight.}
\label{fig.corrected_results}

\end{figure}

We now look at the bottom slice of the model in Figure~\ref{fig.results}(e) in light blue. Looking at the graph, one may observe 2 nodes on the bottom layer that have many red points in the visualization. Each point representing a hole in the layer. This may represent a problem for watertightness, particularly given that this is the bottom layer. Observing the connected components represented by those 2 node in Figure~\ref{fig.results}(e), many holes are visible in the layer due to inadequate resolution of the points. The initial concern about watertightness remains, given that these holes are not covered by a subsequent layer. Finally, the lack of watertightness can be confirmed by looking at the empty space graph in Figure~\ref{fig.software}(d). In this graph, there is a single component representing all empty space. If the model were watertight, at least 2 empty space components would form, one outside the model and one or more inside.

\subsection{Error Corrected Model}

As a comparison, we have computed an error free version of the dragon model. To do this, the triangle mesh provided with the model was subdivided to calculate additional vertices until the point model became watertight. The result of the Mapper and persistent homology calculations can be seen in Figure~\ref{fig.corrected_results}. This new model contained 441,713 points (less than 1\% increase from the original), making it visually indistinguishable from the original.

In Figure~\ref{fig.corrected_results.printed}, the Mapper graph of the filled space looks identical to the Mapper graph of the original in Figure~\ref{fig.results}(a). The persistent homology calculation however is quite different. The number of red dots (i.e.\ holes in the model) have reduced significantly. In fact, the only holes that remain are those representing the major empty cavities of the model's interior.

In Figure~\ref{fig.corrected_results.empty}, the Mapper graph of the empty space is shown. The most important aspect of these new graphs is that there are now 2 connected components. Figure~\ref{fig.corrected_results.empty}(left) represents the connected component of the air surrounding the model. Figure~\ref{fig.corrected_results.empty}(right) represents the air inside the model. The lack of connection between these 2 components indicates that the model is now watertight.

\subsection{Runtime Performance}

\begin{figure}[!h]
\centering

\subfigure[Variable Slice Count: 8-128; Fixed Slice Overlap: 0.05~cm; Fixed Grid Resolution: 0.15~cm\label{fig.performance.slices}]{\includegraphics[width=0.32\textwidth]{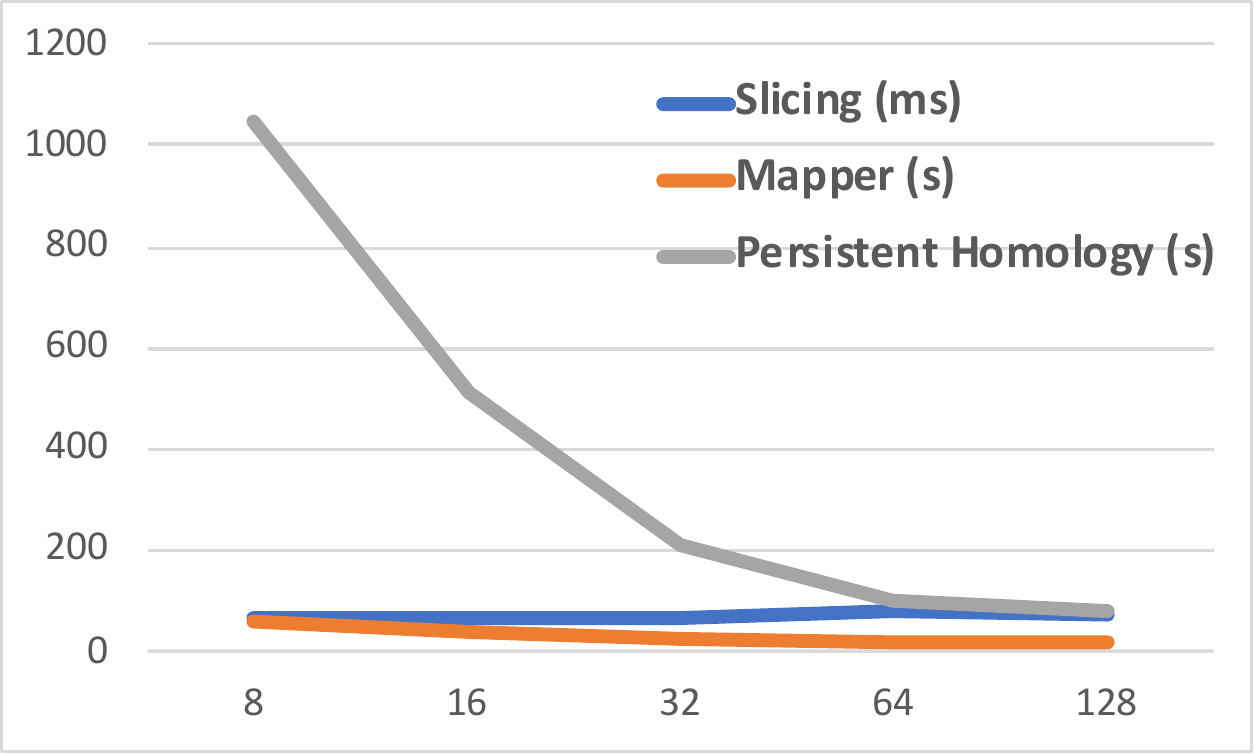}}
\hfill
\subfigure[Fixed Slice Count: 32; Variable Slice Overlap: 0.025-0.2~cm; Fixed Grid Resolution: 0.15~cm\label{fig.performance.overlap}]{\includegraphics[width=0.32\textwidth]{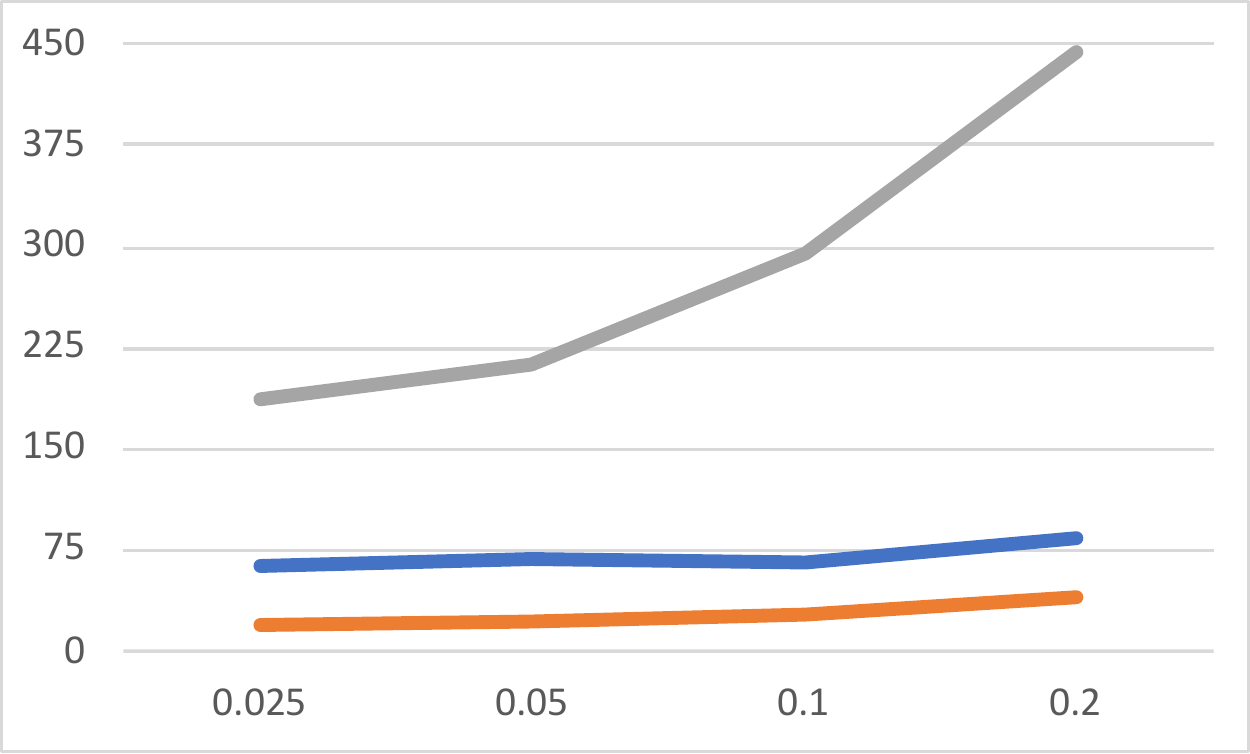}}
\hfill
\subfigure[Fixed Slice Count: 32; Fixed Slice Overlap: 0.1~cm; Variable Grid Resolution: 0.15-0.35~cm\label{fig.performance.grid}]{\includegraphics[width=0.32\textwidth]{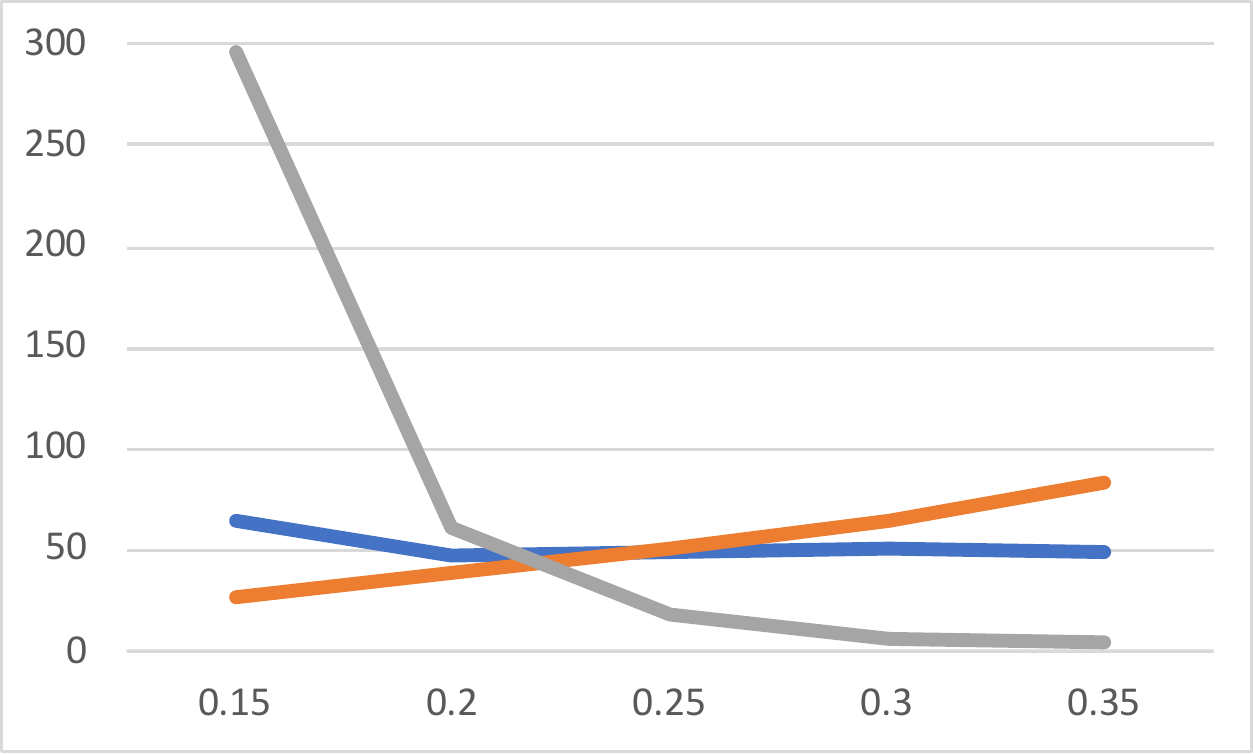}}

\caption{Performance result varying the 3 main parameters of the approach: (a) number of slices, (b) slice overlap, and (c) xy grid resolution. In all results, the time for slicing is presented in milliseconds, while Mapper and persistent homology are reported in seconds.}
\label{fig.performance}

\end{figure}

We tested the runtime performance of our analysis on the Dragon data set by varying the 3 main parameters, the number of slices, slice overlap, and the xy grid resolution. The results can be seen in Figure~\ref{fig.performance}. These results show that persistent homology is almost always the largest cost. This high cost can be attributed to regions that have large connected components. 

\section{Conclusions}
In conclusion, we have presented an approach for using Topological Data Analysis in the evaluation of the quality of 3D printed objects using point cloud-based models. 
We made some simplifying assumptions in this paper. For example, we assume that 3D printing resolution is uniform across the entire xy domain, which is not necessarily true. We also chose a naive rasterization procedure, though any other pre-rasterized model would be adequate for analysis in this pipeline. 

It is also important to note that this approach, as presented, does not report specific problems, aside from watertightness. It instead enables a number of qualitative analyses that depend upon a user's expectation for the output of their model, including certain global or regional problems, such as issues with number of tunnels expected per component; whether the tunnels are connected; the number of connected components per slice; and which connected components make contact slice-to-slice. This essentially enables answering the question, `does the printed model topology match my expectations?'

\section*{ACKNOWLEDGEMENTS}
The dragon model was provided by the Stanford 3D Scanning Repository. This work was supported in part by the National Science Foundation (IIS-1513616).

%\section*{ORCID}
%\orcid{Paul Rosen}{0000-0002-0873-9518}
%\orcid{Junyi Tu}{0000-0001-7026-7454}
%\orcid{Les Piegl}{0000-0003-0629-8496}

% #########################################
% Please pay close attention to the formatting of the references:
%    Authors: Smith, J.; Doe, M.; White, K.:
%    Paper: listed as published
%    Journal: listed in its conventional name
%    Volume: 27(3), 2009, 123-130, i.e. volume, issue, year and pages
%    Conference: name, location, year, pages
%    DOI: https://doi.org/DOI number. You can get the entire link from
%    "https://doi.crossref.org/simpleTextQuery/"
%    Books: author(s), title, publisher, location, year
%    Website: title (if any) and full URL link
% ##########################################

%\referenceSection
%\bibliographystyle{CADA}
\bibliographystyle{abbrv}
\bibliography{refs}

\end{document}